\documentclass[11pt,a4paper,fleqn,epsfig]{article}
\newcommand{\squeezeup}{\vspace{-2.5mm}}
\pdfoutput=1
\usepackage{latexsym}

\usepackage[dvips]{color}
\usepackage{graphicx}

\usepackage{amssymb}
\usepackage{amsmath}
\usepackage{amsfonts}
\usepackage{braket}
\usepackage{units}
\usepackage[left=2.0cm,right=2.0cm,top=3.0cm,bottom=3.0cm]{geometry}

\begin{document}

\title{Quantum Mechanics of a Rotating Billiard}

\author{Nandan Jha$^{1}$ and Sudhir R. Jain$^{2}$\\ 
$^{1}$High Pressure \& Synchrotron Radiation Physics Division,\\
$^{2}$Nuclear Physics Division,\\  Bhabha Atomic Research Centre, Mumbai 400085, India}

\date{}
\maketitle

\begin{abstract}
Integrability of a square billiard is spontaneously broken as it rotates about one of its corners. The system becomes quasi-integrable  where the invariant tori are broken with respect to a certain parameter, $\lambda = 2E/\omega^{2}$ where E is the energy of the particle inside the billiard and $\omega$ is the angular frequency of rotation of billiard. We study the system classically and quantum mechanically in view of obtaining a correspondence in the two descriptions. Classical phase space in Poincar\'{e} surface of section shows transition from regular to chaotic motion as the parameter $\lambda$ is decreased. In the Quantum counterpart, the spectral statistics shows a transition from Poisson to Wigner distribution as the system turns chaotic with decrease in $\lambda$. The wavefunction statistics however show breakdown of time-reversal symmetry as $\lambda$ decreases.
\end{abstract}


\section{Introduction}

A particle moving freely inside a square box and specularly reflecting off the boundary in accordance with the Snell's law defines a square billiard. We consider the situation where the box is rotating about a vertical axis passing through one of its corners while the particle moves freely. The billiard problem thus posed breaks time reversal invariance. This aspect makes the system even more interesting for a quantum mechanical study. The motivation to study this problem in detail is in its intimate, direct relation to the physics of rotating nuclei. Rotating nuclei have been studied in a great detail, beginning from the cranking model by Inglis \cite{inglis}, and the later work by Bohr and Mottelson \cite{bm}. The latter work was brought to conclusion by Jain et al. \cite{srjain1996} where they showed that there were three regimes classified by the solutions of the Duffing equation in the angular momentum space. These results have been subsequently used in understanding several aspects of superdeformed nuclei \cite{review1,review2}.

Classical and quantum billiards with a rotating boundary was first studied by Fairlie and Siegwart \cite{fs1,fs2} and also by Frisk and Arvieu \cite{frisk}. They showed that the phase space portraits exhibited regular to chaotic motion transition with change in perturbation parameter. More studies along these lines were carried out later \cite{borgan}. The billiards considered are related, with respect to the time reversal invariance, to chaotic billiards in the presence of magnetic field. For these systems, the spectral fluctuations belong to the Gaussian Unitary Ensemble of random matrices \cite{berry}. In this paper, we study the classical and quantum signatures of chaos in a square billiard rotating about one of its corners. We study Poincar\'{e} surface of section of classical phase space and show that this system goes from regular to chaotic with change in rotational frequency. Quantum signatures of chaos are studied in the nearest-neighbour level spacing distribution and in the nature of wavefunctions. Wavefunction statistics shows transition from regular to chaotic motion and also reveals the time reversal symmetry breaking.

\section{Classical phase space: Poincar\'{e} surfaces of section}

\begin{figure}[htbp]
\centering
\includegraphics[width=50mm]{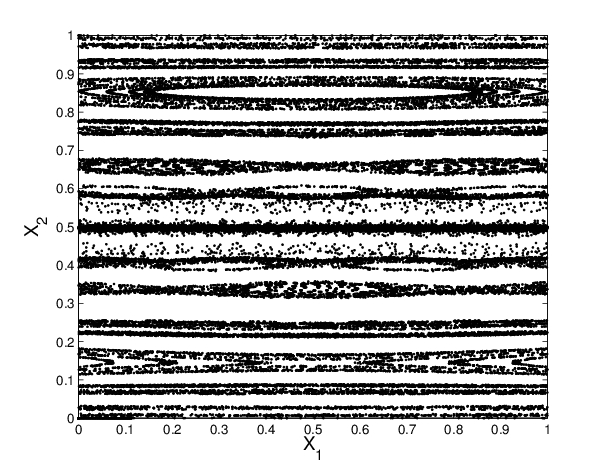}
\includegraphics[width=50mm]{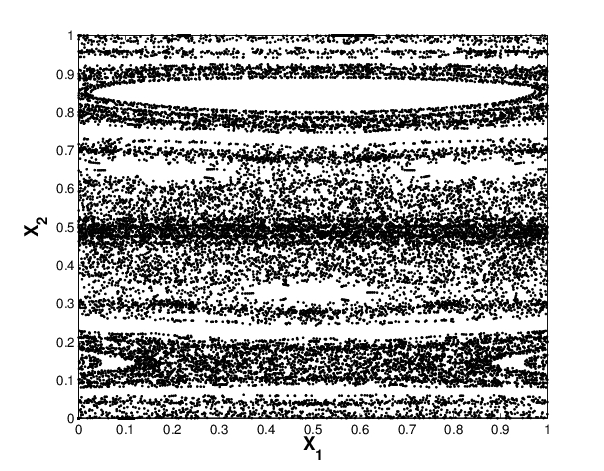}
\includegraphics[width=50mm]{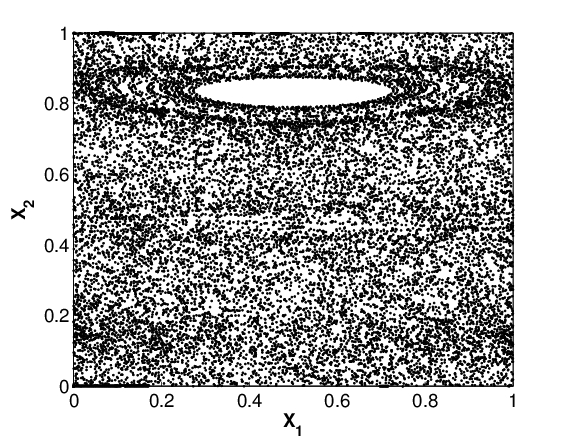}
\caption{Poincare surface section of phase space for $\lambda=10^4$, $\lambda=10^3$ and $\lambda=10^2$. The KAM Tori disappear as $\lambda$ decreases.}
\label{fig:poincare}
\end{figure}

We consider a square billiard rotating about the axis perpendicular to the billiard plane and passing through one of its corners. If we take this corner to be the origin of the coordinate system, then in the rotating frame of the billiard, the Hamiltonian will be
\begin{equation}\label{eq:Ham}
H=\frac{1}{2}\left(p_{x}^{2}+p_{y}^{2}\right)+\omega\left(yp_{x}-xp_{y}\right)+U(x,y)
\end{equation}
$U(x,y)$ is the hard wall potential of the billiard (we consider square billiard with walls at $x=0, x=1, y=0, y=1$) and $\omega$ is the rotational frequency of billiard. The classical dynamics of the system is governed by the parameter $\lambda=2E/\omega^2$ where $E$ is the energy of the particle. For larger values of $\lambda$, the system exhibits more regular behavior. As the value of $\lambda$ decreases, the chaotic region in phase space increases as can be seen in Fig. \ref{fig:poincare}. Here canonically conjugate variables $X_1$ and $X_2$ are plotted for different values of $\lambda$ which are related to the position of the particle and its momentum when it strikes the wall. $X_1$ and $X_2$ are the Birkhoff coordinates on the walls. $X_{1}$ is the fractional distance along the wall the particle strikes increasing in counter-clockwise direction. $X_{2}$ is the normalized canonical conjugate momentum of $X_{1}$.

\noindent\begin{minipage}{.5\linewidth}
\begin{equation*}
X_{1}=\begin{cases}
1-y,& \text{for } x=0\\
y,  & \text{for } x=1\\
x,  & \text{for } y=0\\
1-x,& \text{for } y=1
\end{cases}
\end{equation*}
\end{minipage}%
\begin{minipage}{.5\linewidth}
\begin{equation*}
X_{2}=\begin{cases}
\frac{1}{2}\left(1-\frac{\dot{y}}{\sqrt{2E+\omega^{2}}}\right),& \text{for } x=0\\
\frac{1}{2}\left(1+\frac{\dot{y}}{\sqrt{2E+\omega^{2}}}\right),& \text{for } x=1\\
\frac{1}{2}\left(1+\frac{\dot{x}}{\sqrt{2E+\omega^{2}}}\right),& \text{for } y=0\\
\frac{1}{2}\left(1-\frac{\dot{x}}{\sqrt{2E+\omega^{2}}}\right),& \text{for } y=1
\end{cases}
\end{equation*}
\end{minipage}

\squeezeup

It is seen from phase space plots shown in Fig. \ref{fig:poincare} that with small rotational perturbation, small KAM islands form near the points satisfying nonlinear resonance conditions. As the perturbation increases with decrease in parameter $\lambda$, number of KAM islands decreases and the chaotic area increases. Therefore with change in the value of parameter $\lambda$, the system goes from regular to chaotic with mixed phase space for intermediate values of $\lambda$.

\section{Statistical properties of energy levels}

\begin{figure}[htbp]
\centering
\includegraphics[width=50mm]{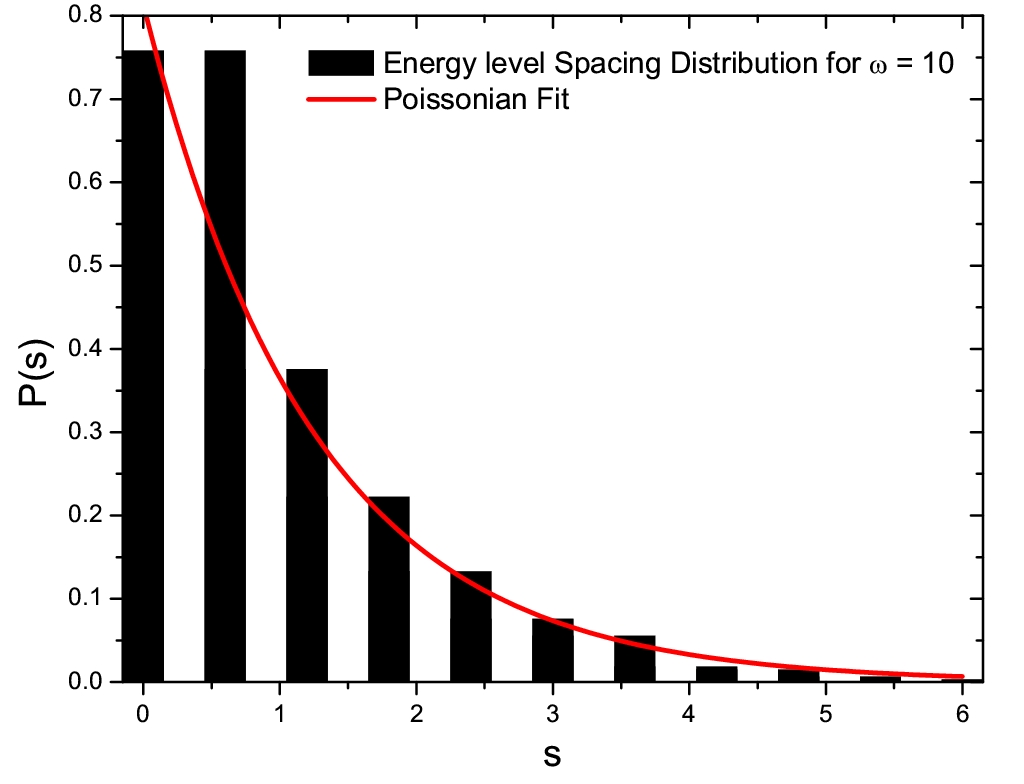}
\includegraphics[width=50mm]{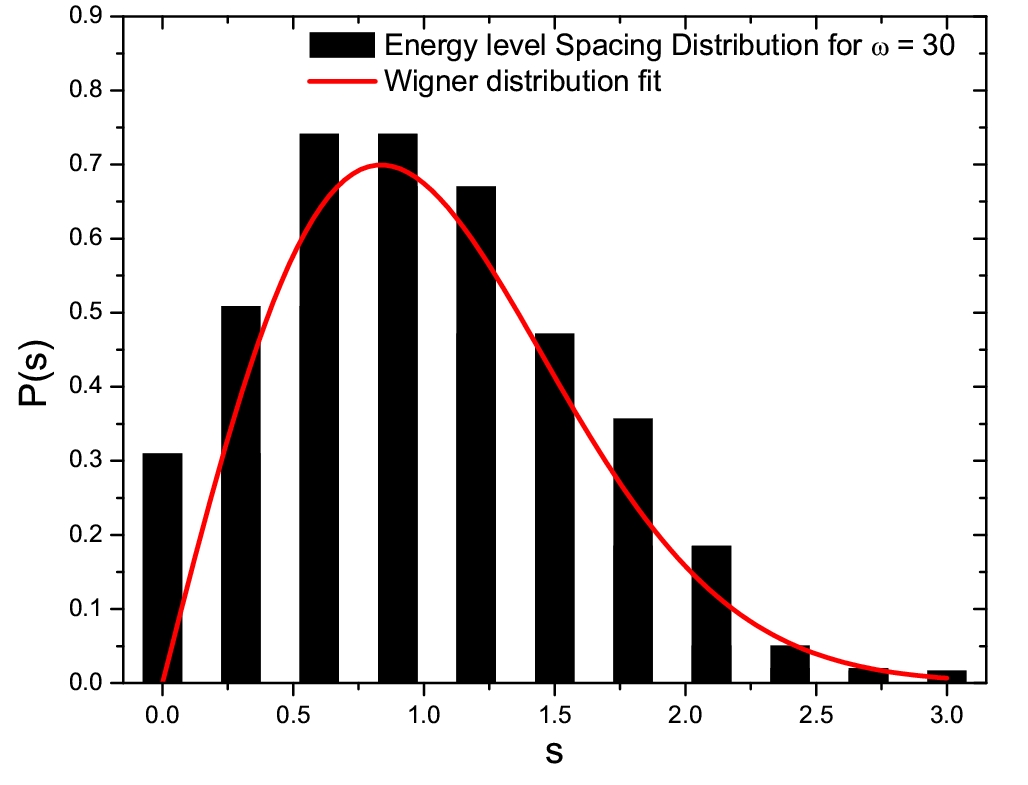}
\includegraphics[width=50mm]{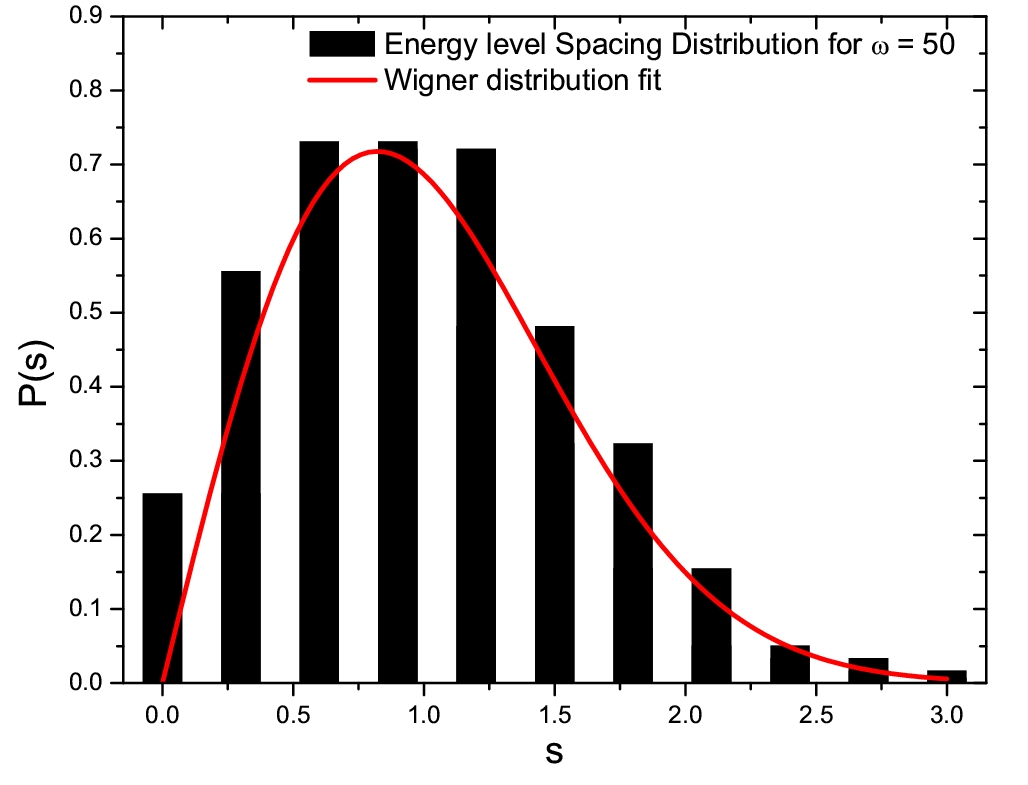}
\caption{Level spacing distribution for $\omega$=10, 30 and 50. The parameter $\lambda$ for the highest energy level ($1000^{th}$ state) is 129, 13.8 and 4.7 for $\omega =$ 10, 30 and 50 respectively. The average value of $\lambda$ for the 1000 levels is 65.2, 6.6 and 2. The distribution changes from Poisson to Wigner as $\lambda$ decreases with increase in $\omega$.}
\label{fig:level_spacing}
\end{figure}

The signature of the onset of chaos can be seen in the nearest neighbor spacing (NNS) distribution calculated for the first 1000 energy levels of the system. These energy levels are calculated by writing the Hamiltonian in Eqn. \ref{eq:Ham} in particle in box wavefunction basis and numerically diagonalizing the resulting Hamiltonian matrix. As we can see from Fig. \ref{fig:level_spacing} that as the value of $\omega$ increases the level spacing distribution changes from Poisson to Wigner like distribution. This is expected from our classical analysis where decrease in $\lambda$ leads to more chaotic motion.

\begin{figure}[htbp]
\centering
\includegraphics[width=60mm]{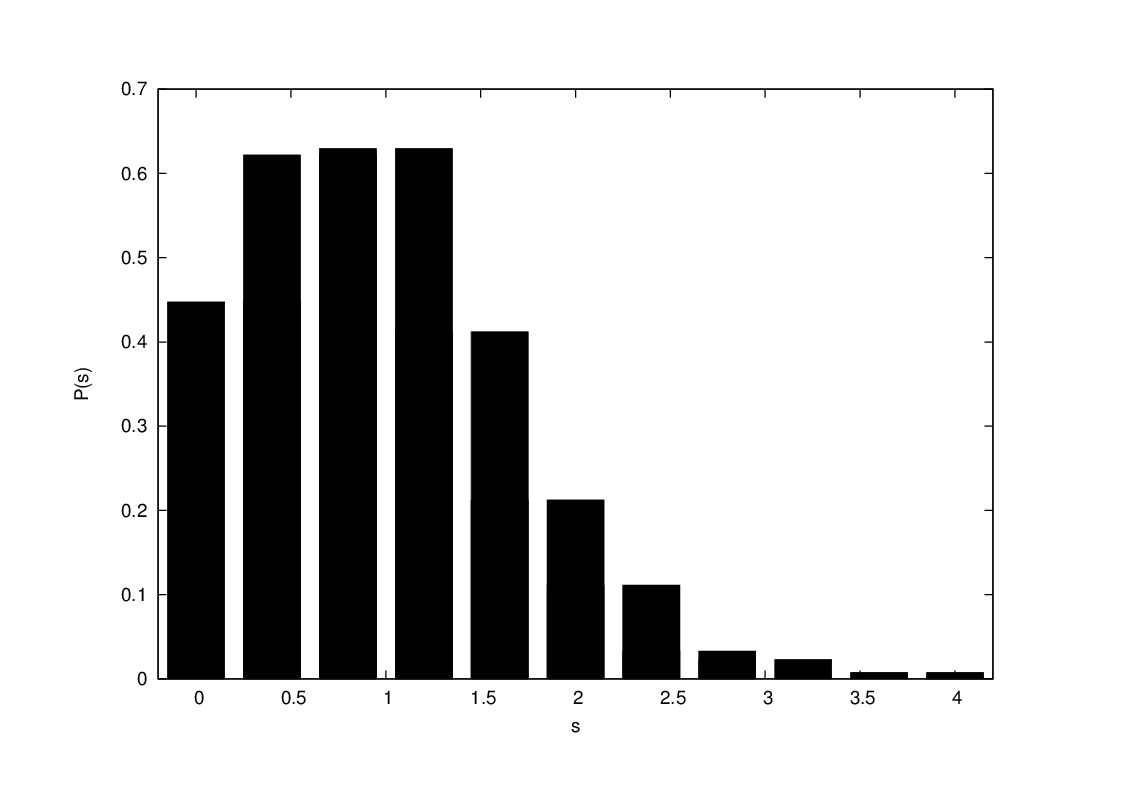}
\includegraphics[width=60mm]{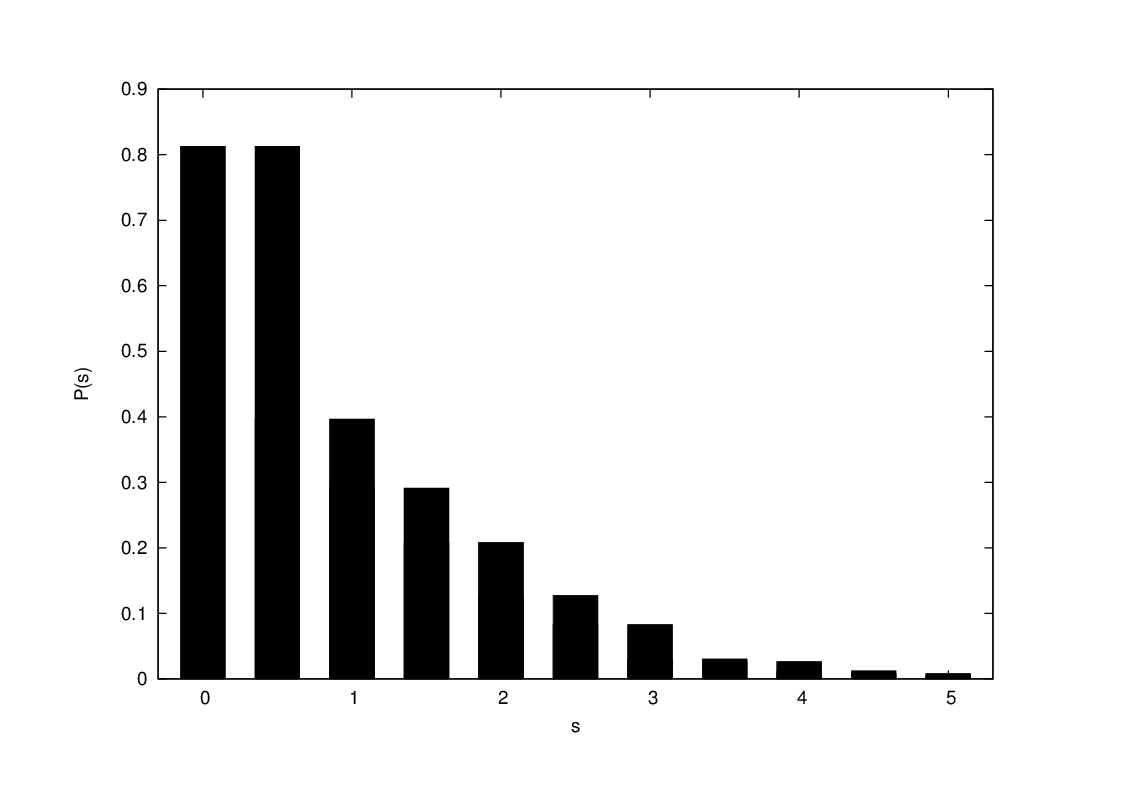}
\caption{Level spacing distribution for 1000 lowest levels (Left) and for 8000th to 9000th levels (Right) at $\omega = 20$.}
\label{fig:high_low_comparison}
\end{figure}

Since the classical dynamics of the system depends on the parameter $\lambda=2E/\omega^2$, therefore we expect that the level spacing distribution will show different dynamics for different energy levels. Therefore, in Fig. \ref{fig:high_low_comparison} we show the level spacing distribution computed from 1000 lowest energy states and for 8000th to 9000th energy levels for $\omega=20$. We can clearly see that for the lower energy levels, the spacing distribution shows departure from Poisson distribution whereas for the higher energy levels, the level spacing distribution is still Poisson. This is because the parameter $\lambda$ for the $1000th$ level is 32 and for the lowest level in the higher energies shown, $\lambda= 251$. Since the energy level sequence of a square box goes to infinity, the level spacing distribution for a very large number of levels will always be Poisson.

\squeezeup
\squeezeup

\section{Wavefunction statistics}

As seen above, the transition from regular to chaotic dynamics is governed by the parameter $\lambda$ which depends on the energy and rotation frequency. This is evident  from the classical phase space as well as from the study of level spacing distribution for lower and higher energy levels at a fixed $\omega$. To get a clearer picture of this transition, wavefunction statistics is important. The reason for the importance for the statistics of wavefunctions is their complexity and analytical intractability. For fully chaotic wavefunctions, it is expected that their amplitude distributions will be well approximated by a normal distribution \cite{berry1}. 

\begin{figure}
\centering
\includegraphics[width=76mm]{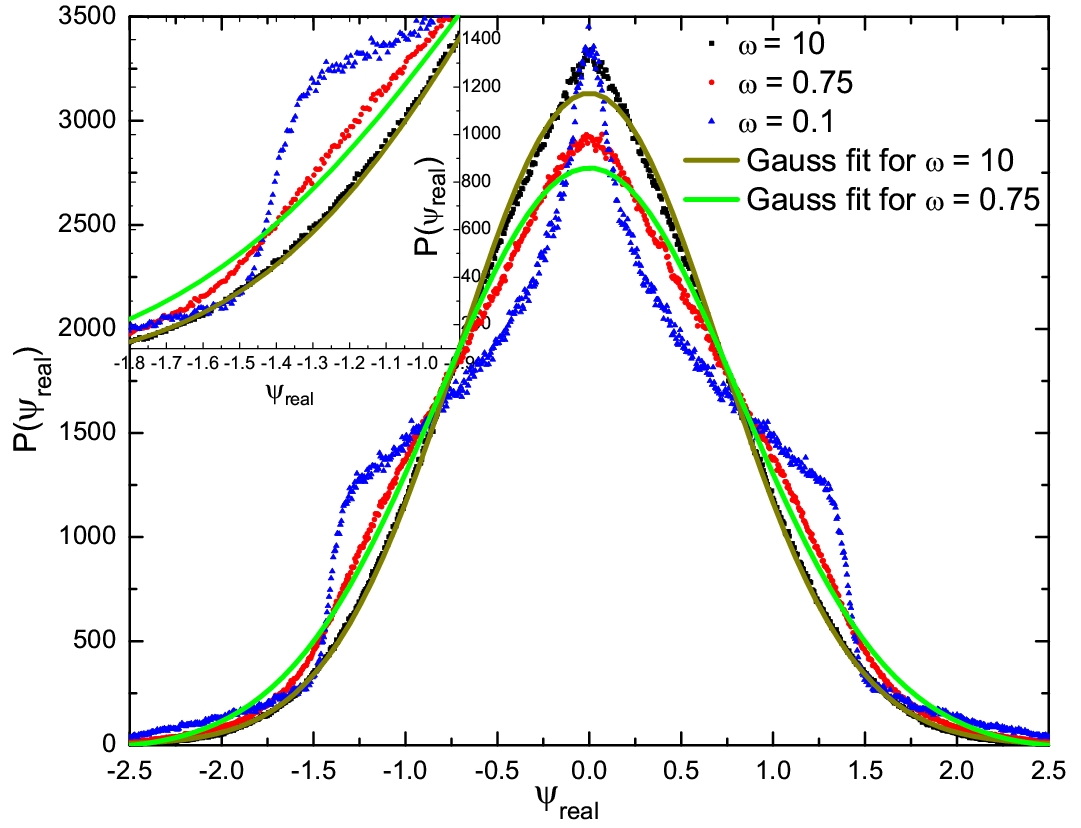}
\includegraphics[width=76mm]{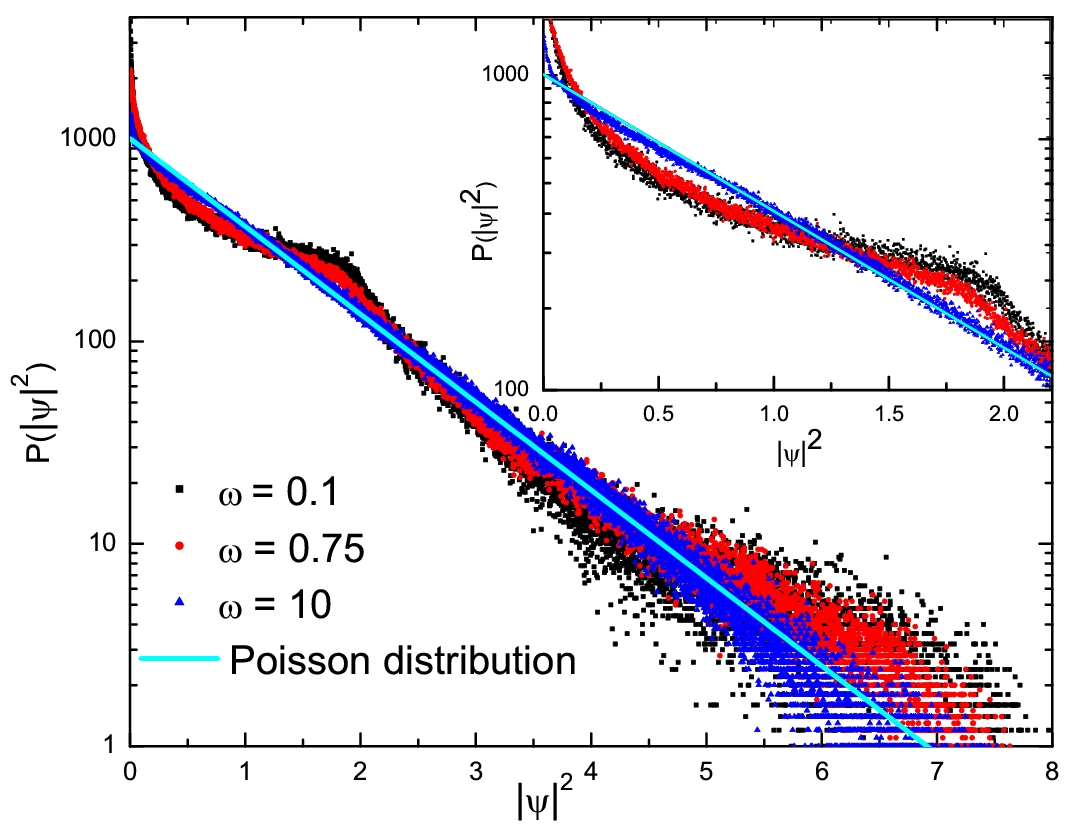}
\caption{Averaged histogram of real part of wavefunctions (left) and of intensity distribution of wavefunctions (right) near $1000^{th}$ wavefunction at different values of $\omega$. The inset in left panel shows the magnified view of the same graph from $\psi=-1.8$ to $-0.9$. The inset in the right graph shows the intensity distribution for $\left|\psi\right|^{2}=$ 0 to 3.}
\label{fig:histogram_of_wavefunction}
\end{figure}

In Fig. \ref{fig:histogram_of_wavefunction}, the probability distribution function of $\psi_{Real}$ and $\left|\psi\right|^{2}$ averaged over ten states near 1000th state are plotted for $\omega=0.1$ $\left(\lambda=1.3\times 10^{6}\right)$, $\omega=0.75$ $\left(\lambda=23102\right)$ and $\omega=10.0$ $\left(\lambda=130\right)$. $\psi_{Real}$ gradually becomes Gaussian distributed and $\left|\psi\right|^{2}$ becomes Poisson distributed as $\omega$ increases from 0.1 to 10.0. Both the statistics of $\psi_{Real}$ and $\left|\psi\right|^{2}$ become chaotic at $\omega=10 \left(\lambda=130\right)$ which is consistent with the classical dynamics shown in Fig. \ref{fig:poincare}. The Poisson distribution of $\left|\psi\right|^{2}$ brings out the possibility of breakdown of time reversal symmetry due to rotation for large $\omega $. One may recall that rotation has similar effect as that of a magnetic field on a charged particle - both correspond to breaking of time reversal. Therefore, the wavefunction statistics shows transition to GUE statistics corresponding to chaotic dynamics with time-reversal symmetry breaking. Another interesting observation is that the probability distribution function of $\left|\psi\right|^{2}$ appears to be a sum of Porter-Thomas like distribution with a very sharp decay near zero and a Poisson like distribution with linear profile in log scale. This may be indication of mixed GUE and GOE like dynamics in the rotating billiard system which needs to be further studied in more detail.

\section{Conclusion}

In this paper, we show that the classical phase space of a particle in a rotating square billiard shows transition from completely regular state to chaotic state as the parameter $\lambda$ decreases. The corresponding behaviour is seen in quantum analogue of this system. As we increase the rotational frequency $\omega$, the level spacing distribution shows transition from Poisson distribution to Wigner like distribution. Wavefunction statistics also captures these features along with a strong indication towards the breakdown of time-reversal symmetry with an increasing $\omega$.

\end{document}